\documentstyle[sprocl,epsf]{article}
\input{psfig}
\def\Journal#1#2#3#4{{#1} {\bf #2}, #3 (#4)}

\def\NIM{\em Nucl. Instrum. Methods}

\def\PRD{{\em Phys. Rev.} D}
\def\ZPC{{\em Z. Phys.} C}
\def\mevc {\ifmmode MeV/{\rm c} \else $MeV/{\rm c}$\fi}
\def\mevcc {\ifmmode MeV/{\rm c}^2 \else $MeV/{\rm c}^2$\fi}
\def\gevc {\ifmmode GeV/{\rm c} \else $GeV/{\rm c}$\fi}
\def\gevcc {\ifmmode GeV/{\rm c}^2 \else $GeV/{\rm c}^2$\fi}

\def\bb{b\overline{b}}
\def\cc{c\overline{c}}

\begin{document}
\title{Time Dependent $B^0 \bar{B}^0$ Mixing at CDF
\footnote{Submitted to the  Proceedings of the 1996 Meeting of the 
Division of Particles and Fields, American Pbysical
Society, Minneapolis MN, August 10-15, 1996.}}
\author{Fritz DeJongh}
\address{Fermilab, PO Box 500 ms 318, Batavia IL 60510}
\author{George Michail}
\address{Harvard University, 44 Oxford St., Cambridge MA 02138}
%
\maketitle\abstracts{
We describe two measurements of $\Delta m_d$.  The first uses
$B \to \nu\ell D^{(*)}$ events and a same-side flavor tagging algorithm.
The second uses dilepton events.  From the average of these two
measurements we find
$\Delta m_d = 0.466 \pm 0.037 \pm 0.031\ \rm ps^{-1}$.
}
\section{Introduction}
Measurements of the frequencies for $B_d$ and $B_s$ mesons to
oscillate into $\bar{B}_d$ and $\bar{B}_s$, respectively,
can potentially constrain the magnitudes of the CKM matrix
elements $V_{ts}$ and $V_{td}$.  These frequencies are proportional
to $\Delta m_d$ and $\Delta m_s$, the mass differences between
the CP eigenstates of the $B_d$ and $B_s$ mesons.
Recent measurements have provided precise determinations of
$\Delta m_d$ and lower limits on $\Delta m_s$~\cite{Paris}.
The large $b\bar{b}$ cross-section
in $p\bar{p}$ collisions at $\sqrt{s}$ $=$ 1.8 GeV has enabled the
reconstruction of large $B$ signals using the CDF detector~\cite{CDF}.
The measurement of a time-dependent mixing probability is made possible
by a precise decay length measurement from the Silicon Vertex Detector
(SVX)~\cite{SVX}. The charges of the decay products tag the flavor of the B
at the time of decay. To tag the flavor of the B at production, several
tagging algorithms have been developed. The measurement of the mistag
probabilities of these algorithms is also useful for future measurements
of CP violation~\cite{Manfred}.

We present herein two measurements of the $B_d$ mixing frequency.
The first uses semileptonic $B$ decays in which the charm has
been fully reconstructed, and a same-side flavor tagging algorithm
using correlations between $B$ mesons and charged tracks.  Such
correlations have been observed at LEP~\cite{LEP} and in 
$B^\pm \to J/\psi K^\pm$
events at CDF~\cite{psik}.
The second uses semileptonic $B$ decays in which the charm has
been inclusively reconstructed, and a flavor tagging algorithm
using the semileptonic decay of the other $B$ in the event.
 
\section{$B^0$ mixing in $B \to \nu\ell D^{(*)}$ events}

For this analysis, we use $B$ mesons reconstructed in the 
following channels:
$\begin{array}{llrl}
B^0  \to \nu\ell^+ D^{*-}, & 
  D^{*-} \to \bar{D}^0 \pi^-_s, & 
    \bar{D}^0 \to &
       K^+ \pi^- \\
 &   & \to & K^+ \pi^-  (\pi^0 \ {\rm not\ reconstructed} )    \\
 &   & \to & K^+ \pi^- \pi^- \pi^+ \\
B^0  \to \nu\ell^+ D^-, & 
  D^- \to K^+\pi^-\pi^-  &  \\
B^+  \to \nu\ell^+ \bar{D}^0, & 
  \bar{D}^0 \to K^+\pi^- & &
      {\rm (Veto \ D^* \ candidates) } \\
\end{array}$

\begin{figure}[h]
\hskip 0.5in
\epsfysize=10.0cm
\epsffile{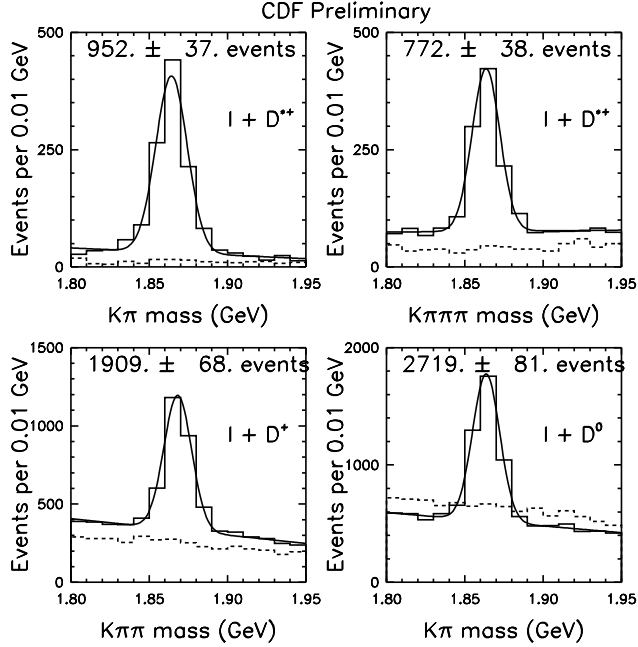}
\caption{Charm signals in semileptonic $B$ decays}
\label{fig:mass4}
\end{figure}

\begin{figure}[h]
\hskip 1.5in
\epsfysize=5.0cm
\epsffile{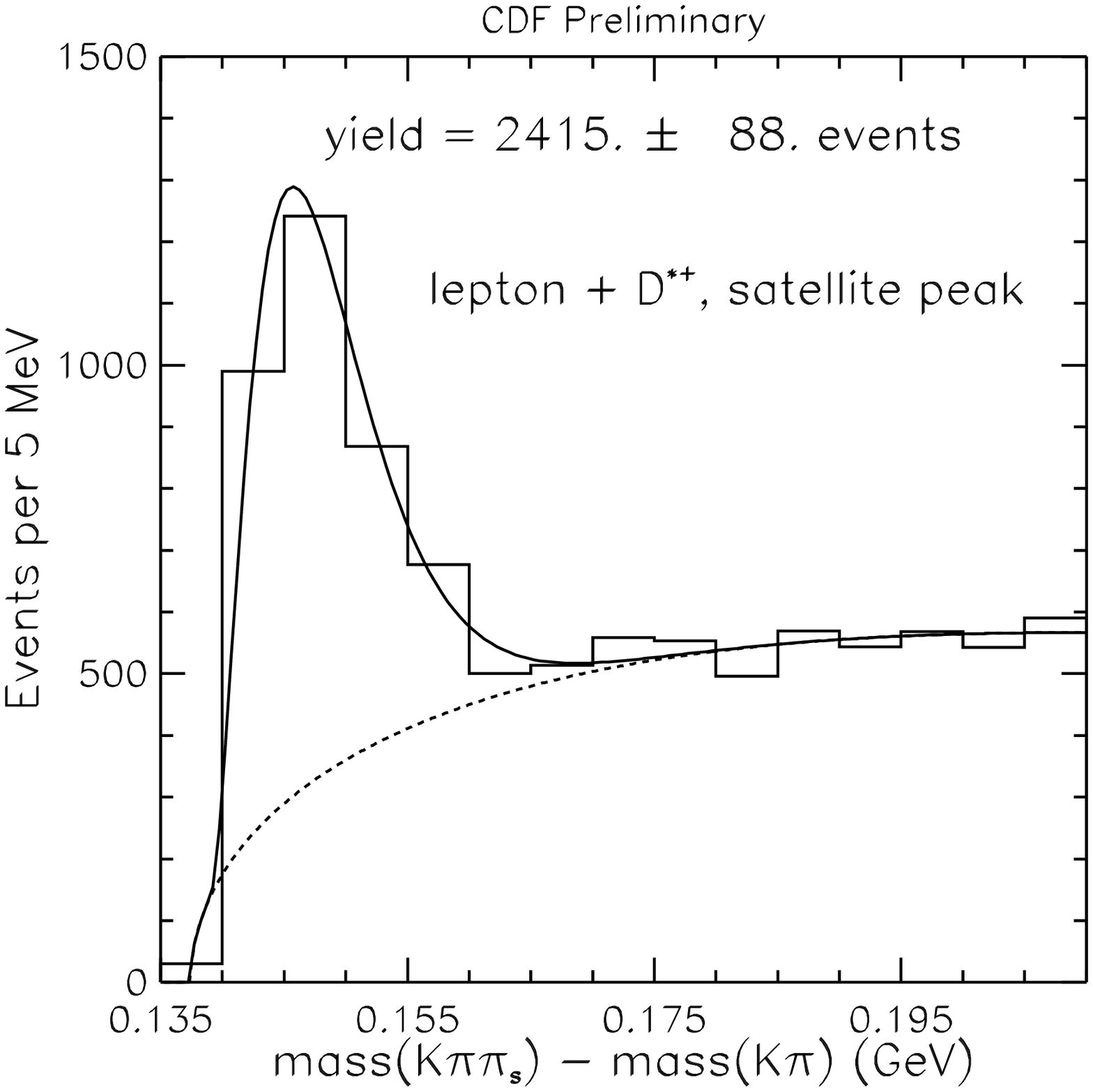}
\caption{$D^*$ signal for 
 $\bar{D}^0 \to K^+ \pi^-  (\pi^0 \ {\rm not\ reconstructed})$ }
\label{fig:kpzs-comb}
\end{figure}

An electron or muon with transverse momentum with respect to the
beam axis ($p_t$) greater than 9 GeV/c triggers the event.  We then
reconstruct the charmed mesons from the tracks in a cone 
of radius 1.0 in $\eta-\phi$ space around the
lepton.  To decrease combinatorial background
from prompt tracks, we select tracks
with impact parameters significantly displaced from the primary interaction
vertex.  The signals are identified as peaks in the mass spectra
of the charm decay products, as shown in Fig.~\ref{fig:mass4} and
Fig.~\ref{fig:kpzs-comb}.

Using the SVX information, we reconstruct the decay length of the $B$
in the plane transverse to the beam axis
 ($L_{xy}^B$).  To obtain the proper decay time
we estimate the boost of the $B$ from the observed decay products and
apply a correction factor for the missing neutrino:
\begin{eqnarray}
c\tau = L_{xy}^B \ \frac{m_B}{p_t(B)} = L_{xy}^B \ \frac{m_B}{p_t(\ell D)} \ K
\end{eqnarray}
On average, we reconstruct 86\% of the momentum of the $B$, 
with an r.m.s. of 11\%.
 
We use a
``Same-side tagging'' (SST) algorithm to tag the flavor of the $B$
at $t=0$.  This algorithm exploits the correlation
between the $B$ flavor and the charge of tracks from either
the fragmentation process or $B^{**}$ decay~\cite{GNR}.
We expect a $B^-$ to be correlated with a $\pi^+$ and a
$\bar{B}^0$ to be correlated with a $\pi^-$.  Due to the production
of $s$ quarks in the fragmentation process, and since we do not
apply $K/\pi$ separation, we expect the observed correlation to
be stronger for the $B^-$ than for the $\bar{B}^0$~\cite{DR}.

For our algorithm, we approximate the $B$ momentum as the
momentum of the reconstructed portion of the $B$.
We define a cone whose axis is the momentum vector
of the $B$, and with radius 0.7 in
$\eta - \phi$ space.  We consider all tracks in this cone with
$p_t$ $>$ 0.4 GeV and which pass within 3 s.d. of the primary vertex.
We define $p_t^{\rm rel}$ for a track as the transverse momentum of the track
relative to the sum of the momenta of the $B$ and that track.
Of the candidate tracks, we select the track with lowest $p_t^{\rm rel}$,
and compare the charge of that track to the charge of the lepton from the
semileptonic decay.  Our efficiency ($\varepsilon$)
for finding such a tag is $\approx 72\%$.

We compare the number of right-sign (RS) correlations
({\it i.e.} $\bar{B}^0 \pi^-$, $B^- \pi^+$) 
to the number of wrong-sign (WS) correlations
({\it i.e.} $\bar{B}^0 \pi^+$, $B^- \pi^-$) 
as a function of $c\tau$.
For the $\bar{B}^0$ we expect the asymmetry $A(t)$:
\begin{eqnarray}
A(t) = \frac{N_{RS}(t) - N_{WS}(t)}{N_{RS}(t) + N_{WS}(t)}
   = D\cos(\Delta m t)
\end{eqnarray}
where $\Delta m$ is the frequency of the oscillation, and
$D$ is the dilution of the flavor tagging algorithm.
$D$ is often expressed in terms of the mistag fraction $w$ as
$D = 1-2w$.
We fit for both $\Delta m$ and $D$.

\begin{figure}[h]
\epsfysize=8.0cm
\hskip 0.5in
\epsffile{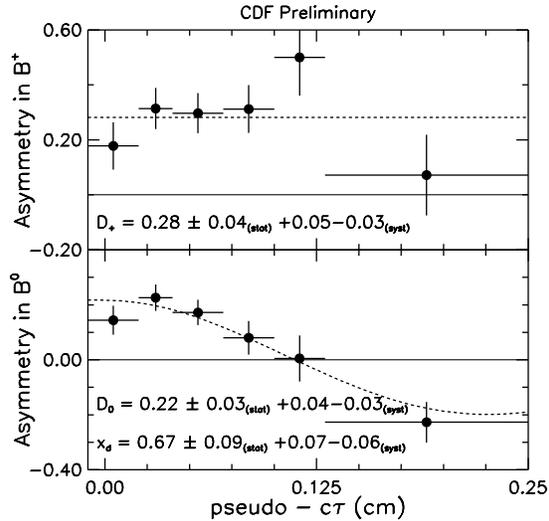}
\caption{Time dependent asymmetry}
\label{fig:result}
\end{figure}

To obtain the asymmetry for $B^0$ or $B^+$, we
correct for the fact that
each signal has contributions from both $B^0$ and $B^+$ decays.
For example, the following decay chains contribute to
the same data sample: \\
$\begin{array}{ll}
B^+  \to \nu\ell^+ \bar{D}^0  & \ {\rm (Veto}\ D^*)       \\
B^0  \to \nu\ell^+ D^{**-} &  D^{**-} \to \bar{D}^0  
( \pi^-_{**}\ {\rm unobserved} )   
\end{array}$ \\
We correct for this cross-talk by performing a fit bin by bin
in $c\tau$.  The inputs to the fit are the raw aymmetries as
measured in each sample for a given $c\tau$ bin, and parameters describing the
$D^{**}$ composition in semileptonic decays.  The outputs are the
true $B^0$ and $B^+$ asymmetries.  
We fit the true $B^0$ asymmetry as a function of $c\tau$ to 
a cosine convoluted with the $c\tau$ resolution function, and extract the
mixing frequency and dilution of the algorithm.
The results are shown in Fig.~\ref{fig:result}.
We also observe an asymmetry for the $B^+$ which is flat with $c\tau$
as expected.

In summary, we find
$\Delta m_d = 0.446 \pm 0.057^{+0.034}_{-0.031}\ \rm ps^{-1}$,
and an effective tagging efficiency for the $\bar{B}^0$,
$\varepsilon D_0^2 = 3.4 \pm 1.0^{+1.2}_{-0.9}\%$.
The dominant systematic uncertainty is from the
fraction of $D^{**}$ in semileptonic $B$ decay.

\section{$B^0$ Mixing in $e\mu$ Events}
For this analysis, we trigger on leptons 
from the semileptonic decay of both $b$ hadrons in an event:
$b_1 \to eX$ and $b_2 \to \mu X$.  We estimate that 70\% of
our signal events come from an $e\mu$ trigger which requires
$p_t(e) > 5\ \gevc$ and $p_t(\mu) > 3\ \gevc$, and 
30\% come from single lepton triggers with
$p_t(\ell) > 9\ \gevc$ and the other lepton found offline.
Offline, we require $M_{e\mu} > 5\ \gevcc$ in order to reject
sequential decays.

The principle of this analysis is to search for an inclusive
secondary vertex associated with one of the leptons.  The decay
length of this vertex and the momenta of tracks associated 
with the lepton provide an estimate of $c\tau$.  
The boost resolution for this technique is
$\approx 21\%$ for the electrons and $\approx 24\%$ for the muons.
The charge of the other lepton provides the flavor tag.

To search for an inclusive secondary vertex, we consider tracks in a 
cone around each lepton that are significantly displaced 
from the primary vertex.  For each lepton we first search for a
secondary vertex with at least two tracks in addition to the lepton with
$p_t$ $>$ 0.5 GeV/c.  If no such vertex is found, we allow
a secondary vertex with only one additional track with
$p_t$ $>$ 1.0 GeV/c.  This algorithm is tuned for high
efficiency near $c\tau = 0$, with the efficiency reaching
a plateau of $\approx 40\%$ for $c\tau > 0.05$ cm according to
a Monte Carlo simulation. 

Since the signal cannot be observed as a narrow peak in a mass
distribution, accounting for backgrounds is a challenge.
We define a fake event as an event with at least one fake lepton.
We have found
that to a very good approximation, the fake electron
events are a subset of the fake muon events, due to the higher electron
$p_t$ cut.  This greatly
simplifies the accounting of fake backgrounds.  To obtain
magnitudes and distributions for fake events, we use the
following samples:
1) Prescaled 5 GeV single electron triggers with another
      track passing all cuts
      except for the presence of a muon stub.
2) $e\mu$ events for which the $\mu$ candidate
      fails quality cuts.  We then assume that fake events for
which the muon passes our selection criteria have the same properties
as these samples.

Other backgrounds arise from sequential decays:  $b\to c\to \ell$.
These backgrounds can be estimated from $p_t^{\rm rel}$ distributions,
and the invariant mass distribution of the secondary vertex tags.
Here, $p_t^{\rm rel}$ is defined as the transverse momentum of the
muon with respect to the highest $p_t$ track in a cone of radius
0.7 in $\eta - \phi$ space around the muon.  We require
$p_t^{\rm rel}$ $>$ 1.25 GeV/c for the muon
in order to reduce sequential backgrounds.
The final sample composition is shown in table~\ref{tab:comp}.

\begin{table}[htb]
\caption{Final sample composition of vertex-tagged $e\mu$ events.
``$e$ tag'' and ``$\mu$ tag'' indicate that the vertex is 
associated with the electron or muon.  The sequential fractions
are fractions of the $b\bar{b}$ component.}
\begin{center}
\begin{tabular}{|l|c|c|} \hline
{\bf Component} & {\bf $e$ Tags} & {\bf $\mu$ Tags} \\ 
\hline \hline 
{Fake e with Real $\mu$}   &  $\leq 1\%$    & $\leq 1\%$  \\ 
{Fake $\mu$ Fraction} & $15\pm 4\%$ & $7\pm 3\%$ \\ \hline 
{$\cc$ events} &  $2\pm 2\%$ &  $4\pm 3\%$ \\ \hline 
{$\bb$ events} & $83\pm 5\%$ & $89\pm 4\%$ \\ \hline
{Sequential $e$ }  &  $8.8\pm 1.3\%$ &  $7.9\pm 1.2\%$ \\
{Sequential $\mu$ }      & $13.6\pm 2.0\%$ & $16.5\pm 2.5\%$ \\ \hline
\end{tabular}
\end{center}  
\label{tab:comp}                         
\end{table}

\begin{figure}[h]
\epsfysize=7.0cm
\hskip 1.0in
\epsffile{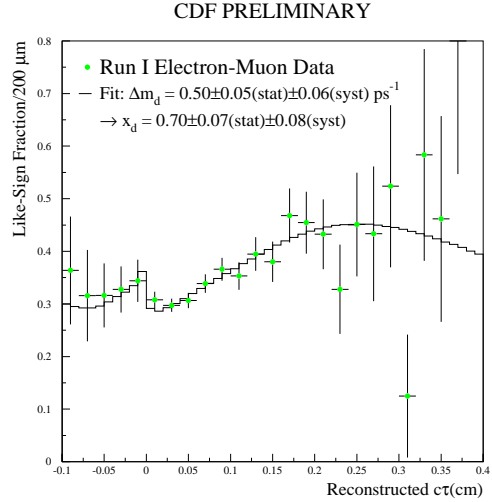}
\caption{Like-sign fraction vs. $c\tau$}
\label{fig:datafit_bl}
\end{figure}

We extract $\Delta m_d$ from a fit to the like-sign fraction as a
function of $c\tau$, with the results shown in Fig.~\ref{fig:datafit_bl}.
This fit includes components for direct $b\bar{b}$, sequential 
$b$ decays,
$c\bar{c}$, and fake events.  
In $\approx 16\%$ of the events with a
secondary vertex found around one lepton, we also find a secondary
vertex around the other lepton.  These events enter the like-sign
fraction plot twice, and we allow for a
statistical correlation between the two entries.
We find
$\Delta m_d = 0.50 \pm 0.05 \pm 0.06\ {\rm ps^{-1}}$, where the 
dominant systematic uncertainties arise from 
uncertainties in the sample composition.

\section{Summary}
We have reported two measurements of $\Delta m_d$.  In
$B\to\nu\ell D^{(*)}$ events 
tagged with a same-side algorithm we find
$\Delta m_d = 0.446 \pm 0.057^{+0.034}_{-0.031}\ \rm ps^{-1}$ and
$\varepsilon D_0^2 = 3.4 \pm 1.0^{+1.2}_{-0.9}\ \%$.
In $e\mu$ events we find  $\Delta m_d = 0.50 \pm 0.05 \pm 0.06\ \rm ps^{-1}$.
The average of these results is 
$\Delta m_d = 0.466 \pm 0.037 \pm 0.031\ \rm ps^{-1}$.

\end{document}